\documentclass[11pt]{article}
\usepackage{a4wide}                      
\usepackage{epsf}                        
\usepackage{latexsym}                    
\usepackage{amsfonts}                    
\usepackage{amssymb}                     

\def\be{\begin{equation}}
\def\ee{\end{equation}}
\def\bea{\begin{eqnarray}}
\def\eea{\end{eqnarray}}
\def\nn{\nonumber \\ [.2cm]}
\def\vsp#1{\vspace{#1}}
\def\hsp#1{\hspace{#1}}

\def\part{\partial}

\def\x{\times}

\def\incl{\mbox{i}}

\def\cR{{\cal R}}

\def\mn{{\mu\nu}}

\def\makeatletter{\catcode`\@=11}
\makeatletter
\def\mathbox#1{\hbox{$\m@th#1$}}%
\def\math@ccstyles#1#2#3#4#5#6#7{{\leavevmode
      \setbox0\mathbox{#6#7}%
      \setbox2\mathbox{#4#5}%
      \dimen@ #3%
      \baselineskip\z@\lineskiplimit#1\lineskip\z@
      \vbox{\ialign{##\crcr
             \hfil \kern #2\box2 \hfil\crcr
             \noalign{\kern\dimen@}%
             \hfil\box0\hfil\crcr}}}}
\def\mathaccstyles{\math@ccstyles\maxdimen}
\def\maththroughstyles{\math@ccstyles{-\maxdimen}}
\def\unity%
 {\maththroughstyles{.45\ht0}\z@\displaystyle {\mathchar"006C}\displaystyle 1}
\begin{document}

\rightline{KUL-TF/04-16}
\rightline{FFUOV-04/15}
\rightline{hep-th/0406148}
\rightline{November 2004}
\vspace{2truecm}

\centerline{\LARGE \bf Giant Gravitons in $AdS_3\times S^3\times T^4$ }
\vspace{.5truecm}
\centerline{\LARGE \bf as Fuzzy Cylinders}
\vspace{1.3truecm}

\centerline{
    {\large \bf B. Janssen${}^{a,}$}\footnote{E-mail address:
                                  {\tt bert.janssen@fys.kuleuven.ac.be}},
    {\large \bf Y. Lozano${}^{b,}$}\footnote{E-mail address:
                                  {\tt yolanda@string1.ciencias.uniovi.es}}
    {\bf and}
    {\large \bf D. Rodr\'{\i}guez-G\'omez${}^{b,}$}\footnote{E-mail address:
                                  {\tt diego@fisi35.ciencias.uniovi.es}}
                                                            }

\vspace{.4cm}
\centerline{{\it ${}^a$ Instituut voor Theoretische Fysica, K.U. Leuven}}
\centerline{{\it Celestijnenlaan 200D,  B-3001 Leuven, Belgium}}

\vspace{.4cm}
\centerline{{\it ${}^b$Departamento de F{\'\i}sica,  Universidad de Oviedo,}}
\centerline{{\it Avda.~Calvo Sotelo 18, 33007 Oviedo, Spain}}

\vspace{2truecm}

\centerline{\bf ABSTRACT}
\vspace{.5truecm}

\noindent
Using the non-Abelian action for coincident Type IIB gravitational waves
proposed in hep-th/0303183 we show that giant gravitons in the
$AdS_3\times S^3\times T^4$ background can be described in terms of
coincident waves expanding into a fuzzy cylinder, spanned by two embedding scalars and 
one worldvolume scalar.  This fuzzy cylinder has dipole and magnetic moments
with respect to the 2-form and 6-form potentials of the background, and can be
interpreted as a bound state of D1-branes and D5-branes (wrapped on the 4-torus) wrapped
around the basis of the cylinder. We show the exact agreement between this description
and the Abelian, macroscopical description given in the literature.

\newpage
\section{Introduction}

There is strong evidence by now that giant gravitons are described microscopically
in terms of dielectric gravitational waves, expanding into massless higher-dimensional $p$-branes.
{}From this point of view, the expansion of the waves happens because the transverse coordinates to the
coincident waves are matrix valued, which allows the waves to couple non-trivially to the RR
potentials of the background. The effect is analogous to Myers dielectric (or magnetic
moment) effect for D-branes \cite{Myers}.

The non-Abelian worldvolume effective action for
coincident D$p$-branes contains non-Abelian couplings to RR potentials of order higher than $(p+1)$.
Under the influence of a RR $(p+4)$-field strength the stable configuration
corresponds to an expansion of the D$p$-branes into a D$(p+2)$-brane with
topology $R^{1,p}\times S^2_{\rm fuzzy}$. This brane is stable because it carries
a dipole (or magnetic) moment with respect to the RR potential, that cancels the
contraction due to its tension.

Stable expanded brane configurations had been previously found
in the literature \cite{Emparan} as single spherical D$(p+2)$-branes with
non-vanishing dipole moment and D$p$-brane charge dissolved in their worldvolumes.
In fact, this is the large $N$ limit of the $N$ D$p$-branes expanding into a fuzzy
D$(p+2)$-brane \cite{Myers}. The description of expanded brane configurations
in terms of the expanding non-Abelian D$p$-branes is commonly referred in the 
literature as the microscopical description, whereas the description in terms of 
the spherical Abelian D$(p+2)$-brane is referred as the macroscopical description. 
A non-trivial check of the validity of a given microscopical configuration of branes
is its agreement with the corresponding macroscopical description when the number of
branes is large.

Giant gravitons in $AdS_m\times S^n$ spacetimes have been studied macroscopically in
\cite{GST,GMT,HHI,DJM,DJM2} as stable brane configurations with non-zero angular
momentum (or graviton charge) wrapped around $(m-2)$- or $(n-2)$-spheres in the
spacetime background and with a non-vanishing dipole or magnetic moment with respect to
the background gauge potential. The dynamical equilibrium of these configurations is reached
through the cancellation between the tension of the brane and the coupling of the
angular momentum to the background flux field. At this equilibrium point, the configurations
can be shown to have a zero mass, and behave essentially as massless, finite size objects, which
explains the name of ``giant gravitons''.

Giant gravitons expanded in the spherical part of $AdS_m\times S^n$
were first considered \cite{GST}
as a possible way to realise the stringly exclusion principle \cite{MS}.
The radius of these expanded gravitons is proportional to their
angular momentum, and since this radius is bounded by the radius
of the $S^n$, the configuration has associated a maximum angular momentum, in
agreement with the CFT predictions. Giant gravitons expanding in the $AdS$ part of the geometry
do not satisfy however the stringy exclusion principle \cite{GMT,HHI},
given that $AdS$ is non-compact. They are referred in the literature as ``dual giant gravitons''
(whereas we will refer to the giant gravitons in $S^n$ as ``genuine giant gravitons''). 
Some discussion about how the stringy
exclusion principle can still be satisfied having these two types of giant graviton configurations
can be found in \cite{GMT,DJM} and in \cite{BS,Berens}.

In the microscopic picture, the giant gravitons are generated by gravitational waves expanding
into the spherical brane configurations due to Myers dielectric effect \cite{DTV,BMN, JL1}.
Therefore, in order to describe giant gravitons microscopically we need a non-Abelian action
for coincident gravitational waves,
and then to identify the dielectric couplings that will be
responsible for their expansion.
The non-Abelian effective action describing coincident
gravitational waves in arbitrary backgrounds was constructed in \cite{JL2, JLR}\footnote{
Giant gravitons in plane wave backgrounds have also been studied microscopically
in \cite{BMN,Mik,DMS,SY,TT,CS}.}.
Using this action 
we provided a microscopical description for the giant gravitons in
$AdS_m\times S^n$ spacetimes expanding into 2-spheres:
the genuine giant graviton in $AdS_7\times S^4$, the
dual giant graviton in $AdS_4\times S^7$ and the genuine and dual giant gravitons
in $AdS_5\times S^5$.\footnote{The giant and dual giant gravitons in $AdS_5\times S^5$ involve a
fuzzy 3-sphere which is however described in terms of an Abelian $S^1$ bundle over a fuzzy
2-sphere. See \cite{JLR} for the details of this construction.}
Other macroscopical giant graviton solutions are known in these spacetimes,
though the microscopic discussion is much harder, mainly due to the technical difficulties
involved in the construction of fuzzy $n$-spheres with $n>2$ \cite{Ramgoolam}.
For example, the genuine giant
graviton of
$AdS_4\times S^7$ and the dual one of $AdS_7\times S^4$ involve the construction of a fuzzy 5-sphere.
The microscopical description of these gravitons has not been
given yet, though we hope to report on this in a forthcoming paper \cite{JLR2}.

In this paper we center on the microscopical description of giant gravitons in another background,
with its own pecularities, namely $AdS_3\times S^3\times T^4$. It is known that macroscopic giant
gravitons in $AdS_3\times S^3\times T^4$ have features which are different from the ones in
$AdS_m\times S^n$ spacetimes with $m, n > 3$. First of all, giant gravitons in
$AdS_3\times S^3\times T^4$ only exist when their angular momentum has a  very specific value, namely
a multiple of the number of branes that create the geometry. Moreover, for this value of the
momentum the graviton can have arbitrary size. The fact that the potential governing the size of
the giant graviton is flat in this background was already noted in \cite{GST,HHI}. Therefore, these
configurations seem to be completely unrelated to the stringy exclusion principle in this background
\cite{MS}. In fact, it has been suggested in \cite{MM,LMS} that giant gravitons are not the
correct supergravity description of the chiral primary states of the D1-D5 system, which are in
turn described by the more general family of metrics given in \cite{BBKR,MM, LM}.

Secondly, it is also known that various types of macroscopic
giant graviton configurations can exist in this
background. Besides the usual distinction between genuine and dual giant gravitons, living
respectively in the spherical and the $AdS$ part of the spacetime, it is possible to construct
so-called mixed giant gravitons, which are basically a linear combination of the previous two.
The one-cycle these mixed giant gravitons are wrapped on is the sum of the one-cycles in the
$AdS$ and in the $S$ parts. Also these configurations can have arbitrary size in either part of
the geometry, for the specific value of the angular momentum mentioned above. Furthermore, all
types of giant gravitons (genuine, dual or mixed) can be built by using  D1-branes, D5-branes or both. 
The D1's wrap the one-cycle in the adequate part of the background, while the D5's wrap the 
same one-cycle and the $T^4$ part.\footnote{It should be clear that the D5 giant gravitons are in
fact the $T$-duals of the D1's, after dualisation over the directions of the $T^4$.} In this paper
we will directly deal with the most general case: combined D1-D5, mixed giant
gravitons, with the understanding that each separate case (or combinations thereof) can be
obtained by putting the appropriate parameters to zero.

A specific problem of the microscopic giant gravitons (of all types) in $AdS_3\times S^3\times T^4$
seems to be the fact that the gravitational waves should expand into a fuzzy $S^1$.\footnote{For those 
blowing up into a D1 giant graviton, or a fuzzy $S^1$ times an Abelian $T^4$ for the D5 giant 
gravitons.} The question then  arises how such a fuzzy $S^1$ can be realised. The solution proposed 
here is that the waves expand into a fuzzy cylinder, spanned by the $S^1$ in the background geometry 
and a worldvolume scalar field $\omega$.\footnote{The fuzzy cylinder has also been shown to play a 
role \cite{BL, Bena} in the microscopical
description of the supertube \cite{MT}.}
This scalar field arises naturally from T-duality in the 
derivation of the Type IIB wave action \cite{JL1}.
Since the worldvolume field $\omega$ has no geometrical meaning in the background, one effectively
sees the cylinder for a given value of this scalar field, which results into an $S^1$.

This paper is organised as follows. In section 2 we discuss briefly the $AdS_3\times S^3\times T^4$
background and how it arises as the near horizon limit of a D1-D5 intersection. Its main purpose is
to set our notation. In section 3 we discuss macroscopically the most general case
of combined D1-D5, mixed giant gravitons living both in $AdS_3$ and $S^3$. In section 4 we
construct the microscopic picture, deriving first the non-Abelian action for gravitational waves in
Type IIB, involving the scalar field needed for the cylinder algebra. We also discuss some properties
of the fuzzy cylinder and its algebra and we come to the actual construction of the microscopic
picture in section 4.3. We summarize our conclusions in section 5.

\section{The background}

The $AdS_3\times S^3\times T^4$ background arises as the near horizon geometry
of the intersecting D1-D5 system \cite{Malda, MS}
\bea
ds^2 &=& H_1^{-1/2} H_5^{-1/2} (-dt^2 + dz^2)
   +  H_1^{1/2} H_5^{1/2}  (d\rho^2 + \rho^2 d\Omega_3^2) + H_1^{1/2} H_5^{-1/2}  dy_m^2, \nn
e^{\Phi} &=&  H_1^{1/2} H_5^{-1/2},  \hsp{1.5cm}
F_{tz\rho} = \part_\rho H_1^{-1}, \hsp{1.5cm}
F_{\alpha_1  \alpha_2 \alpha_3} =  - \rho^3 \part_\rho H_5 \sqrt{g_{S^3}},
\eea
where the $\alpha_i$ are the angles and $g_{S^3}$ the determinant of the metric on the $S^3$ in the
overall transverse space. The coordinates $y_m$ $(m=1,...,4)$, describing the relative transverse
space,
can in principle have an unlimited range, though here we will choose them periodic with period $2 \pi$.
This can be thought of as wrapping the D5 on a four-torus.\footnote{In general the D5 can be
wrapped on any Ricci-flat surface $M^4$, giving rise in the near horizon limit to spacetimes of the form
$AdS_3 \x S^3 \x M^4$.} The harmonic functions $H_1$ and $H_5$ are given by
\be
H_1 = 1 + \frac{Q_1}{\rho^2}, \hsp{2cm}
H_5 = 1 + \frac{Q_5}{\rho^2},
\ee
with $Q_1$ and $Q_5$ the total D1- and D5-brane charge. In the near horizon limit $\rho \rightarrow 0$, 
this solution goes to $AdS_3\times S^3\times T^4$
\bea
ds^2 &=& \frac{\rho^2}{L^2} (-dt^2 + dz^2) + \frac{L^2}{\rho^2}d\rho^2 + L^2 d\Omega_3^2
+ {\cal R}^2  dy_m^2  \nn
e^{\Phi} &=& {\cal R}^2, \hsp{1.5cm}
F_{tz\rho} = \frac{2\rho}{Q_1}, \hsp{1.5cm}
F_{\alpha_1  \alpha_2 \alpha_3} = 2 Q_5 \sqrt{g_{S^3}}\, ,
\eea
where the radius $\cR$ of the 4-torus as well as the radii $L$ of curvature of $AdS_3$ and $S^3$,
which coincide in this background, are functions of the numbers of D1-branes and
D5-branes:
\begin{equation}
{\cal R}^2=\sqrt{{Q_1}/{Q_5}}, \hsp{1cm}   L^2=\sqrt{Q_1Q_5} .
\end{equation}

In this paper, we will work in global coordinates for $AdS_3$:
\begin{eqnarray}
ds^2 &=&-\Bigl(1+\frac{r^2}{L^2}\Bigr) dt^2+ \Bigl(1+\frac{r^2}{L^2}\Bigr)^{-1} dr^2 + r^2 d\varphi^2 \\
     && \hsp{2cm}   + L^2(d\theta ^2 + \cos^2\theta d\phi^2 + \sin^2\theta d\chi^2)
                    + {\cal R}^2  dy_m^2,\nn
e^{\Phi} &=& {\cal R}^2, \hsp{1.5cm}
C^{(2)}_{t\varphi} = -\frac{Q_5}{L^3}r^2, \hsp{1cm}
C^{(2)}_{\phi\chi} = Q_5\sin^2\theta\, , \label{2form}
\end{eqnarray}
\noindent where
$\varphi, \phi, \chi, y_m \in [0,2\pi]$ and $\theta\in [0,\pi]$.  The RR background field $C^{(2)}$ can
also be expressed in terms of its Hodge-dual 6-form potential as:
\begin{equation}
\label{6form}
C^{(6)}_{\phi\chi 1234}=Q_1\sin^2\theta\, ,\qquad
C^{(6)}_{t\varphi 1234}=-\frac{Q_1}{L^3}r^2\, .
\end{equation}
Notice that the same expressions for the 6-form potential arise from performing four T-duality
transformations along the $T^4$ directions, after also renaming $Q_1$ and $Q_5$. It is clear that an
$AdS_3\times S^3\times T^4$ solution, due to the number of isometry directions, can be embedded both in
Type IIA as Type IIB supergravity. The above form, however, with all four radii of the $T^4$ being
equal, is only a solution of Type IIB. We will limit ourselves for the rest of this paper to
this specific case.

\section{Macroscopic giant gravitons in $AdS_3\times S^3\times T^4$}

As we have mentioned, the most general giant graviton solution
in the $AdS_3\times S^3\times T^4$ background
is in terms of a test brane consisting on a bound state of D1-branes
and D5-branes wrapped on the 4-torus, both with angular momentum in $S^3$
and expanding at the same time in the $AdS_3$ and in the $S^3$ parts of the
geometry, while maintaining constant radii in these spaces \cite{LMS}.
Therefore we look at a trial solution with
$\theta={\rm constant}$, $r={\rm constant}$ and $\phi=\phi(\tau)$, where
$t=\tau$ in static gauge, that is, the giant graviton runs around the sphere along
the coordinate $\phi$. We wrap our combination of D1-branes and wrapped D5-branes around
the circle parametrised by\footnote{One could either choose wrapping the branes
around the circle parametrised by $(\varphi-\chi)/2$, in which case the graviton
moves along $\phi$ in the opposite direction. This is a consequence of the symmetry of
the background under the simultaneous interchange $\phi\rightarrow -\phi$,
$\chi\rightarrow -\chi$.}
\begin{equation}
\psi=\frac{\varphi+\chi}{2}\, .  
\ee
Then the pullbacked metric on the D-branes is given by
\begin{equation}
\label{mixedline}
ds^2=-(1+\frac{r^2}{L^2})dt^2+(r^2+L^2\sin^2{\theta}) d\psi^2+L^2\cos^2{\theta}
d\phi^2 + \cR^2  dy_m^2\, .
\end{equation}
We have as well non-vanishing, constant RR potentials given by
(\ref{2form}) and (\ref{6form}). 

Considering a giant graviton made of $n$ D1-branes and $m$ D5-branes\footnote{Alternatively
one can consider $m$ D5-branes with $n$ D1-brane charge dissolved in their worldvolumes,
and arrive at the same results.}
we obtain the
following action, after integrating over $\psi$:
\begin{eqnarray}
\label{macaction}
S&=&-2\pi\Bigl(nT_1+(2\pi)^4 m T_5 \frac{Q_1}{Q_5}\Bigr)
\int d\tau \left[\sqrt{\frac{Q_5}{Q_1}} \sqrt{(r^2+L^2\sin^2{\theta})
               \Bigl (1+\frac{r^2}{L^2}-L^2\cos^2{\theta}{\dot{\phi}}^2 \Bigr)} \right.\nonumber\\
&& \left.  \hsp{7cm}
-Q_5 \Bigl(\frac{r^2}{L^3}+\sin^2{\theta}\dot{\phi}\Bigr)\right]
\end{eqnarray}
with $T_1$ ($T_5$) the tension of a D1-brane (D5-brane). For the Hamiltonian
we get
\begin{eqnarray}
\label{Hmac}
H&=&\frac{2\pi T_1 (nQ_5+m Q_1)}{L}\Bigl[\frac{1}{\cos{\theta}}
\sqrt{1+\frac{r^2}{L^2}}\nonumber\\
&&\sqrt{\cos^2{\theta}(\sin^2{\theta}+\frac{r^2}{L^2})+
\Bigl( \frac{P_\phi}{2\pi T_1 (nQ_5 +m Q_1)}-\sin^2{\theta}
\Bigr)^2}-\frac{r^2}{L^2}\Bigr]\, ,
\end{eqnarray}
where we have taken into account that $T_1=(2\pi)^4 T_5$.
$P_\phi$ is the angular momentum carried by the combination of branes,
which is constant given that $\phi$ is a cyclic coordinate in (\ref{macaction}).
It is clear that the minimum energy solution corresponds to
\begin{equation}
\label{mom}
P_\phi=2\pi T_1 (nQ_5+mQ_1)\, ,
\end{equation}
for which
\begin{equation}
H=\frac{2\pi T_1 (nQ_5+mQ_1)}{L}=\frac{P_\phi}{L}\, ,
\end{equation}
and this happens independently of the
size of the giant graviton. It is easy to see that the genuine and dual
giant graviton solutions given in \cite{HHI} arise as the special limits
$r=0$ or $\theta=0$ of this solution, since in these limits the giant graviton
expands either on the $S^3$ or on the $AdS_3$ part of the geometry.
That the potential governing the size of the
giant graviton is flat in the $AdS_3\times S^3$ background was already noted in
\cite{GST,HHI}. This fact poses a puzzle with the realisation of the stringy
exclusion principle. However, as we mentioned in the introduction, giant graviton
configurations do not seem to be the correct supergravity description
of the chiral primary states of the
dual CFT \cite{MM,LMS}. Indeed
in the CFT there are no chiral primary states beyond
$J_L=J_R=Q_1Q_5$ \cite{MS}, whereas giant gravitons
only exist with angular momentum $P_\phi=2\pi T_1 (nQ_5+mQ_1)$, a result which
seems to be completely unrelated to the CFT predictions.

\section{The microscopical description}

We expect to describe microscopically the giant graviton configuration
of the previous section
in terms of coincident gravitons expanding into a 1-brane with the topology
of a ``fuzzy'' circle, consisting
on a bound state of D1-branes and D5-branes wrapped on the 4-torus.
In this description the expansion of the gravitons takes place due to their interaction
with the RR 2-form and 6-form potentials of the background, which in this case have both electric
and magnetic components. At the level of the graviton worldvolume effective action
this interaction occurs in the form of non-Abelian dielectric and magnetic
moment couplings.

\subsection{The action}

The effective action describing a system of coincident gravitons
in Type IIB was constructed in \cite{JL1} for weakly curved backgrounds, and later
extended to more general ones in \cite{JLR}. The extended action presented in 
\cite{JLR} can be used to study the $AdS_m\times S^n$ background, which is not a 
linear perturbation to Minkowski. This action was truncated for
simplicity to certain worldvolume fields equal to zero. One of these fields will
however be non-vanishing in the $AdS_3\times S^3\times T^4$ background, so our first task
will be to extend the computation in \cite{JLR} to this case.

The worldvolume dynamics of Type IIB gravitons is determined by D-strings and
F-strings ``ending'' on them \cite{JL1}, in a way which is manifestly S-duality
invariant. These strings are wrapped around a transverse direction that appears 
automatically as an isometric direction in the T-duality derivation of the action. 
This direction is in fact the direction along which the T-duality is performed.
Let us call it the $z$-direction, and $l^\mu$ the Killing vector pointing along 
$z$, i.e. $l^\mu=\delta^\mu_z$ in the adapted coordinate system. Each type of 
string ``ending'' on the gravitons has associated a worldvolume scalar forming 
an invariant field strength either with $\incl_l C^{(2)}$ or with $\incl_l B^{(2)}$, 
where $\incl_l$ denotes the interior product with $l^\mu$.
We call $\omega$ the worldvolume scalar associated to F-strings.
This worldvolume scalar plays the role of the T-dual of the $z$-direction.
Since we are dealing with a gauged sigma model in which the translations along
this direction are gauged, the embedding scalar 
$Z$ disappears as a transverse scalar but a new worldvolume scalar $\omega$ is generated 
in the process which accounts for the corresponding degree of freedom.
This situation is analogous to the one that is found in the relation between the
NS5-brane and the Kaluza-Klein monopole via T-duality. In this case the transverse
direction in which the NS5-brane is dualised becomes the Taub-NUT direction of the
monopole, which is isometric in the action, and a new worldvolume scalar is generated
which is associated to wrapped F-strings ending on the monopole \cite{EJL}.

The effective action for Type IIB gravitons constructed in \cite{JLR} is modified for 
non-vanishing, but constant $\omega$ in the following way (we restrict for simplicity 
to vanishing RR 4-form potential):
\begin{equation}
\label{action}
S_{\rm W_B}^{\rm BI}=-T_0\int d\tau {\rm STr}\Bigl\{ k^{-1}
\sqrt{-P[E_{\mn}+E_{\mu i}(Q^{-1}-\delta)^i_k E^{kj}E_{j\nu}]
{\rm det}(Q^i_j)}\Bigr\}\, ,
\end{equation}
where now
\begin{eqnarray}
\label{Qs}
&&E_{ij}={\cal G}_{ij}\, ,  \hspace{2cm}
E_{iz}=e^{\Phi}k^{-1}l^{-1} (i_k C^{(2)})_i\, ,
\nn
&&E_{zi}=-E_{iz}\, ,   \hspace{1.6cm}
E_{zz}=l^{-2}\, , \nn
&&Q^i_j=\delta^i_j \unity
+ie^{-\Phi}kl[X^i,X^k]{\cal G}_{kj}-i[X^i,\omega] (i_kC^{(2)})_j \, , \nn
&&Q^i_z=ie^{-\Phi}kl^{-1}[X^i,\omega]+i[X^i,X^k](i_kC^{(2)})_k \, , \nn
&&Q^z_i=ie^{-\Phi}kl[\omega,X^k]{\cal G}_{ki}\, , \nn
&&Q^z_z=\unity+i[\omega,X^k](\incl_k C^{(2)})_k \, .
\end{eqnarray}
Here $i,j,k$ exclude the $z$-direction and ${\cal G}_{\mu\nu}$ denotes
the reduced metric, typical for gauged sigma models
\be
{\cal G}_{\mu\nu}=g_{\mu\nu}-k^{-2}k_{\mu}k_{\nu}-l^{-2}l_{\mu}l_{\nu},
\ee
which projects out the embedding scalars corresponding to the two isometry 
directions $l^\mu$ and 
$k^\mu$, where the latter is pointing along the direction of propagation of
the gravitons (see below).
The scalar $k$ and vector $k_\mu$ are defined as  $k^2=g_{\mu\nu}k^\mu k^\nu$ and 
$k_\mu=g_{\mu\nu}k^\mu$. The same notation applies to $l^\mu$.  We have also taken in the above action 
that $g_{\mu\nu}k^\mu l^\nu=0$, a condition that is satisfied for the background
that we consider in this paper.

The Chern-Simons part of the action contains the term \cite{JL1}
\begin{equation}
\label{actionCS}
S_{\rm W_B}^{CS}=-i T_0\int d\tau \ {\rm STr}\{ {P[(\incl_{[X,\omega]})C^{(2)}]} \}
                = -i T_0\int d\tau \ {\rm STr}\{ [X^i,\omega] C^{(2)}_{ij}DX^j \}
\end{equation}
which will be playing an important role in the description of the giant
graviton, as we will see.

The fact that the direction of propagation
appears as an isometric direction is common to all gravitons in Type II and
M theories (see \cite{JL2, JLR}). It is in fact easy to see that, in the
Abelian limit, a Legendre transformation restoring the dependence along the time
derivative of this direction yields the usual action for massless particles in
terms of an auxiliary $\gamma$-metric (see the previous references for the details).
The second isometric direction, $z$, is however special to the Type IIB case.
The reason why the dependence along the T-duality direction cannot be restored in
this case seems to be a technical one. It is remarkable however that only due to
the presence of this isometric direction we can obtain the right dielectric
couplings to higher order Type IIB RR potentials relevant in the background we are 
considering.

\subsection{The fuzzy cylinder}

We expect to describe microscopically the giant graviton configuration
of section 3 in terms
of gravitons expanding into a
1-brane with the topology of a ``fuzzy'' circle with radius
(see (\ref{mixedline}))
\begin{equation}
R=\sqrt{r^2+L^2\sin^2{\theta}}\, .
\end{equation}
It is clear however that a circle cannot be made non-commutative unless we
embed it in a higher dimensional non-commutative manifold. The simplest thing
is to embed it in a non-commutative cylinder.

A non-commutative version of the circle condition $x_1^2+x_2^2=R^2$
can be obtained by making the
non-commutative Ansatz
\begin{equation}
\label{fuzzycyl}
[X^1,X^2]=0\, ,\qquad [X^1,X^3]=ifX^2\, ,\qquad [X^2,X^3]=-ifX^1
\end{equation}
i.e. taking the coordinates $X^1$ and $X^2$, defining the circle, 
together with a third generator $X^3$, to satisfy the algebra of the 
two-dimensional Euclidean group, which is the algebra defining the fuzzy cylinder
\cite{BL,CDP,Hyak,Hashi}, with $X^1$, $X^2$ parametrising the base and
$X^3$ the axis of the cylinder. The length scale $f$ is the non-commutative
parameter.

The quadratic Casimir associated to the algebra (\ref{fuzzycyl})
is $(X^1)^2+(X^2)^2$,
so the base of the cylinder is indeed a non-commutative circle, since we can
realise the condition $x_1^2+x_2^2=R^2$ as
\begin{equation}
(X^1)^2+(X^2)^2=R^2\unity\, .
\end{equation}
But, which generator in our geometry is associated to $X^3$? It turns out
that the worldvolume scalar $\omega$, that couples in the effective action
describing the system of coincident gravitons (\ref{action}) and (\ref{actionCS}),
must play the role of the coordinate $X^3$ along the axis of the cylinder. Therefore the
fuzzy cylinder is not a geometrical object in which the gravitons expand, given that
it is not defined enterely in terms of embedding scalars with the interpretation of 
transverse coordinates. This will become clearer below when we construct the giant 
graviton solution explicitly.

The representations of the fuzzy cylinder algebra are infinite
dimensional (see for instance \cite{BL}).
This means that we can only provide a microscopical
description of the giant graviton solution for an infinite number
of gravitons. Still, even though the dimension of the matrices is infinite, 
the algebra is non-trivial, since the non-commutative parameter $f$ is 
independent of the dimension of the representation.
This situation is different from the fuzzy $S^2$ case, where the limit of
infinite number of gravitons is at the same time the commutative limit.

One explicit realisation of (\ref{fuzzycyl}) is \cite{Hyak,Hashi}
\begin{equation}
\label{coords}
X^1=\frac12 \rho_c T^1\, ,\qquad X^2=\frac12 \rho_c T^2\, ,\qquad \omega=-f T^3\, ,
\end{equation}
where we already
take $\omega$ as the direction along the axis, with
\begin{eqnarray}
(T^1)_{mn}&=&\delta_{m+1,n}+\delta_{m-1,n} \nonumber\\
(T^2)_{mn}&=&i\delta_{m+1,n}-i\delta_{m-1,n} \nonumber\\
(T^3)_{mn}&=&(m-\frac12)\delta_{m,n}
\end{eqnarray}
and $m$ and $n$ running from $-\infty$ to $+\infty$. The quadratic Casimir depends
on the parameter $\rho_c$ as
\begin{equation}
\label{casimircyl}
(X^1)^2+(X^2)^2=\rho_c^2\unity\, ,
\end{equation}
so we can have a fuzzy version of the circle defined by $x_1^2+x_2^2=R^2$ if we choose
$\rho_c=R$ and we embed the circle in a cylinder whose axis is taken along the $\omega$ 
direction and is therefore infinite, since the eigenvalues of $\omega$ range from 
$-\infty$ to $+\infty$.

The length of the cylinder is in fact related to the non-commutative parameter $f$.
The matrix algebra (\ref{fuzzycyl}) has a discrete translational symmetry
along $\omega$, with shift unit $f$ \cite{Hashi}. The unitary matrix $U$
defined by $U_{mn}=\delta_{m+1,n}$ acts as
\begin{equation}
\label{transl}
U^{-1}X^i U=X^i\, ,\, i=1,2\, ; \qquad U^{-1}\omega U=\omega +f\, ,
\end{equation}
so the fuzzy cylinder is invariant under translations along $\omega$ with no
deformation in the $(X^1,X^2)$-plane, with shift unit
fixed by the non-commutative parameter $f$. This parameter can then be regarded as a minimal
distance in the $\omega$-direction, and the size of this direction can
be estimated as
\begin{equation}
\label{long}
l=f{\rm Tr}\unity\, ,
\end{equation}
which is indeed infinite for $f\neq 0$.\footnote{It is however possible to take the
commutative limit such that the resulting cylinder has finite length by taking $f$ 
going to zero exactly to compensante the divergence of ${\rm Tr}\unity$.}

In the commutative limit $f\rightarrow 0$ the system is invariant under continuous
translations along $\omega$. This reflects a symmetry under overall translations
in the direction of the axis of the cylinder, which is analogous to that of the supertube
\cite{MT,BHO}.
We must stress however that in our case the worldvolume scalar $\omega$ does not
have an interpretation as a transverse coordinate,\footnote{Given that $\partial \omega=0$ we
can neither interpret it as inducing F-string charge in the configuration (recall that $\omega$ 
forms an invariant field strength with $i_l B^{(2)}$, ${\cal F}=\partial\omega + i_l B^{(2)}$ 
\cite{JL1}). If that were the case we would be describing a configuration different from 
the giant graviton that we want to study.} so the circle that we are making
non-commutative by embedding it in the cylinder is not
physically located in a cylinder in the transverse space. On the contrary,
in our construction one effectively sees the cylinder for a given value of $\omega$,
and this results into an $S^1$.
{}From this point of view it is natural to find that
the fuzzy cylinder has an invariance along this direction.

\subsection{The microscopic giant graviton}

Looking at the background metric given by (\ref{mixedline}) we expect the gravitons to
expand into a fuzzy circle with radius $R=\sqrt{r^2+L^2\sin^2{\theta}}$.
In the most general case this fuzzy circle corresponds to a bound state of $n$ D1-branes
and $m$ D5-branes wrapped on the 4-torus.

Taking Cartesian coordinates
\begin{equation}
x_1=\sqrt{r^2+L^2\sin^2{\theta}}\cos{\psi}\, ,\qquad
x_2=\sqrt{r^2+L^2\sin^2{\theta}}\sin{\psi}
\end{equation}
the line effective element (\ref{mixedline}) as seen by the waves reduces to $(i=1,2)$
\begin{equation}
ds^2=-(1+\frac{r^2}{L^2})dt^2+ dx_i^2 + 
L^2\cos^2{\theta}d\phi^2+{\cal R}^2 dy_m^2 \, ,
\end{equation}
and the electric and magnetic 2-form and 6-form potentials to
\begin{equation}
C^{(2)}_{\phi i}=-Q_5 \frac{\sin^2{\theta}}{r^2+L^2\sin^2{\theta}}
\epsilon_{ij}x_j\, ,
\qquad
C^{(2)}_{ti}=\frac{Q_5}{L^3}\frac{r^2}{r^2+L^2\sin^2{\theta}}
\epsilon_{ij}x_j\, ,
\end{equation}
\begin{equation}
C^{(6)}_{\phi i 1234}=-Q_1 \frac{\sin^2{\theta}}{r^2+L^2\sin^2{\theta}}
\epsilon_{ij}x_j\, ,
\qquad
C^{(6)}_{ti1234}=\frac{Q_1}{L^3}\frac{r^2}{r^2+L^2\sin^2{\theta}}
\epsilon_{ij}x_j\, .
\end{equation}

We now make the fuzzy cylinder Ansatz for the 2-sphere parametrised by $x^1,x^2$
and the worldvolume scalar $\omega$ appearing in the wave action:
\begin{equation}
[X^1,X^2]=0\, ,\qquad [X^1,\omega]=if X^2\, ,\qquad [X^2,\omega]=-if X^1\, .
\end{equation}
Rewriting the 6-form potentials in terms of their hodge dual, the Ansatz for the R-R potentials becomes
\begin{eqnarray}
C^{(2)}_{\phi i}&=&-(nQ_5+mQ_1)\frac{\sin^2{\theta}}{r^2+L^2\sin^2{\theta}}\epsilon_{ij}X_j\, ,
\nonumber\\
C^{(2)}_{ti}&=&\frac{(nQ_5+mQ_1)}{L^3}\frac{r^2}{r^2+L^2\sin^2{\theta}}\epsilon_{ij}X_j\, .
\end{eqnarray}
Here, $C^{(2)}_{\phi i}$ couples in the BI part of the non-Abelian wave action (see ({\ref{Qs})),
and $C^{(2)}_{ti}$ in the CS part, given by (\ref{actionCS}).

Taking into account that the gravitons propagate along the $\phi$-direction,
so that $k^\mu=\delta^\mu_\phi$, and substituting the background and
the non-commutative Ansatz above in the
action for coincident gravitons (\ref{action}), (\ref{actionCS}),
we find:
\begin{eqnarray}
S_{\rm W_B}&=&-T_0 \int d\tau {\rm STr}\Bigl\{\frac{1}{L\cos{\theta}}
\sqrt{1+\frac{r^2}{L^2}}\nonumber\\
&&\sqrt{\Bigl(\unity -f(nQ_5+mQ_1) \frac{\sin^2{\theta}}{r^2+L^2\sin^2{\theta}}
(X^i)^2\Bigr)^2+\frac{f^2}{L^2} (nQ_5+mQ_1)^2 \cos^2{\theta}(X^i)^2}\nonumber\\
&&-f(nQ_5+mQ_1)\frac{1}{L^3}\frac{r^2}{r^2+L^2\sin^2{\theta}}(X^i)^2\Bigr\}\, .
\end{eqnarray}
In this description, since the direction of propagation is isometric, we are
effectively dealing with a static configuration, and we can compute the potential
as minus the Lagrangian. Note that the embedding scalars $X^i$ only appear via their 
quadratic Casimir (\ref{casimircyl}). Also, it is easy to check that the corrections due to 
the contributions of $(X^i)^{2n}$ in the symmetrised trace prescription vanish, so that 
we can write the potential exactly in terms of an ordinary trace. We stress that
this is an exact expression, in contrast
to the giant gravitons expanded in fuzzy 2-spheres \cite{JL2,JLR},
where the symmetrised trace induces corrections of order $1/N^2$.
We find a potential:
\begin{eqnarray}
\label{micpot}
V_{\rm W_B}&=& \frac{{\rm Tr}\unity T_0}{L}f (nQ_5+mQ_1)
\Bigl[\frac{1}{\cos{\theta}}
\sqrt{1+\frac{r^2}{L^2}} \nonumber\\
&&\sqrt{\cos^2{\theta}(\sin^2{\theta}+\frac{r^2}{L^2})+
\Bigl(\frac{1}{f(nQ_5+mQ_1)}-\sin^2{\theta}\Bigr)^2}-\frac{r^2}{L^2}\Bigr]\, .
\end{eqnarray}
Here ${\rm Tr}\unity$ is infinite, since the irreducible representations of the
fuzzy cylinder algebra are infinite dimensional.
The potential per unit length of the cylinder is however finite, and it is given by
\begin{eqnarray}
\label{micpot2}
{\cal V}_{\rm W_B}&=& \frac{T_0}{L} (nQ_5+mQ_1)
\Bigl[\frac{1}{\cos{\theta}}
\sqrt{1+\frac{r^2}{L^2}} \nonumber\\
&&\sqrt{\cos^2{\theta}(\sin^2{\theta}+\frac{r^2}{L^2})+
\Bigl(\frac{1}{f(nQ_5+mQ_1)}-\sin^2{\theta}\Bigr)^2}-\frac{r^2}{L^2}\Bigr]\, ,
\end{eqnarray}
where we have used (\ref{long}).

The minimum energy is reached when the non-commutative parameter
\begin{equation}
\label{fnc}
f=(nQ_5+mQ_1)^{-1}\, ,
\end{equation}
for which the radius of the cylinder remains however arbitrary. For this value of $f$ the
energy per unit length is
\begin{equation}
{\cal E}=\frac{T_0}{L}(nQ_5+mQ_1)=\frac{P_\phi}{L}\, ,
\end{equation}
with $P_\phi$ the momentum (\ref{mom}). Therefore, the configuration describes a massless
brane with momentum $P_\phi$, and we find perfect agreement with the
macroscopical description.

We can compare the microscopical potential
(\ref{micpot2}) and the minimum energy condition (\ref{fnc})
with the corresponding quantities in the macroscopical calculation. Indeed,
the momentum
and energy that should be compared to those of the macroscopical calculation
are the ones per unit length of the cylinder, since we need to project onto the
$(X^1,X^2)$-plane to make connection with the description in terms of the string
wrapped around the circle $x_1^2+x_2^2=R^2$. The corresponding quantities in the
macroscopical calculation are
given
by (\ref{Hmac}) and (\ref{mom}), respectively. We then find that there is exact agreement,
given that the macroscopical momentum $P_\phi$ is given, in microscopical quantities, by
\begin{equation}
\label{identification}
P_\phi=\frac{{\rm Tr}\unity T_0}{l}=\frac{T_0}{f}\, ,
\end{equation}
where we have taken into account that microscopically the momentum is given by the
tension of a wave times the number of them, and we have used (\ref{long}).
We must stress that, as shown in
\cite{Myers}, the agreement between the microscopical
or non-Abelian calculation and the macroscopical or Abelian one is found when
the number of expanding branes goes to infinity. For the giant
graviton configurations that we have studied microscopically in \cite{JL2,JLR},
which involved fuzzy 2-spheres,
we indeed found this agreement for infinite number of gravitons. In the case of the
$AdS_3\times S^3\times T^4$ background the irreducible
representations of the fuzzy cylinder are
infinite dimensional, so our calculation is only valid for an infinite number of
gravitons. From this point of view it is no surprise that we find exact agreement
with the macroscopical description.

The crucial difference between the fuzzy cylinder
and the fuzzy $S^2$ is that for the fuzzy $S^2$
the limit of infinite number of gravitons is at the same time the commutative limit,
so the microscopical description, in terms of a fuzzy $S^2$, tends in this limit to the
macroscopical, or commutative, description, formulated in terms of a classical
spherical test brane. In the case of the fuzzy
cylinder the non-commutative parameter $f$ is independent of the dimension of the
representation and can therefore be independently sent to zero. This justifies why
the agreement between the microscopical and macroscopical descriptions
occurs for $f\neq 0$, and therefore for a non-commutative
cylinder. Physically this is due to the fact that the gravitons do not expand
onto the whole cylinder, but only onto its projection on the $(X^1,X^2)$-plane.
Therefore, the value of the shift unit along the axis of the cylinder, which is
given by $f$, as discussed around (\ref{transl}), should be physically irrelevant.
In the classical limit, however, when
the integer charges of the D1-branes and D5-branes that create the background
geometry, $Q_1$ and $Q_5$, are very large, so that the supergravity solution is
valid, $f$, as given by (\ref{fnc}), must be very small, so in this
limit the cylinder indeed becomes commutative.

\section{Conclusions}

We have shown that the action proposed in \cite{JLR} to describe multiple Type
IIB gravitational waves is suitable for the microscopical study
of giant gravitons in the $AdS_3\times S^3\times T^4$ background
in terms of dielectric gravitational waves.
This action was used in \cite{JLR} to describe the giant gravitons in the
$AdS_5\times S^5$ background as expanded gravitational waves. The genuine (dual) giant 
graviton was shown to be described in terms of gravitational waves expanding into a fuzzy
3-sphere contained inside $S^5$ ($AdS_5$),
with a non-vanishing magnetic (dipole) moment with respect to the RR 4-form
potential of the background. In both cases for large
number of gravitons we found perfect agreement with the macroscopical
description of \cite{GST,GMT,HHI}, which provided a strong support for the validity of
our action. Remarkably, the fuzzy $S^3$ solution to the equations of motion derived
from our action could simply be described as an $S^1$ bundle over a fuzzy $S^2$ base
manifold.

In this paper we have used the same action to describe microscopically the giant
graviton configurations of another Type IIB background, $AdS_3\times S^3\times T^4$.
This background has the special feature that both giant and dual
giant gravitons can be studied in a unified way in terms of strings winding at the
same time around a circle in $AdS_3$ and in $S^3$ \cite{LMS}. This can happen because
the background has both electric and magnetic RR 2-form potentials switched on.
That genuine giant gravitons
(satisfying stringy exclusion principle for general $AdS_m\times S^n$
backgrounds) and dual giant gravitons
(not satisfying stringy exclusion principle) can be described in a
unified way in this background is in consonance with the fact that giant graviton
configurations in the
$AdS_3\times S^3\times T^4$ background do not seem to be the supergravity duals
of the chiral primary states of the two dimensional CFT of the D1-D5 system. Instead,
they seem to be associated to ``dissasociated'' D1-D5 systems \cite{LMS}.
Therefore in this spacetime giant graviton configurations would be completely
unrelated to the stringy exclusion principle.

It would perhaps shed some light on this issue to study dual descriptions of IIB waves
expanded onto circular D1- and (wrapped) D5-branes along the lines of \cite{BS},
where the relations between giant and dual giant graviton configurations in M-theory
and Polchinski and Strassler's \cite{PS} $N D3\rightarrow D5$ and 
$N D3\rightarrow NS5$ dielectric brane configurations are used to argue that each
type of giant graviton can only exist in a given regime of the space of 
parameters\footnote{We
would like to thank a suggestion of the referee along these lines.}.

We should stress that one limitation of our microscopical description is
that we are constrained to work with infinite coincident gravitons, given that the
representations of the algebra of the fuzzy cylinder are infinite dimensional.
A possible way to consider finite $N$ coincident gravitons would be by taking the
fuzzy cylinder algebra as the limit of a fuzzy ellipsoid, for which the algebra
is a deformed $SU(2)$ algebra (see for instance \cite{Hashi}). In this case
however a circular section as the one involved in the giant graviton configurations
can only be recovered in the limit in which the ellipsoid becomes a cylinder, for which
we would be stuck again with a infinite number of gravitons.

Giant gravitons in $AdS_m\times S^n$ spacetimes
preserve the same fraction of the supersymmetries than the point-like
gravitons in these spacetimes. The condition over the spinors reads \cite{GMT,HHI}:
\begin{equation}
(\Gamma^{t\phi}+1)\epsilon=0\, ,
\end{equation}
which coincides with the supersymmetry preserving condition of a gravitational wave
with momentum $P_\phi$ propagating in flat space \cite{BKO}. One would expect that
the same holds true in the $AdS_3\times S^3\times T^4$ background, although this has
not been checked out explicitly. On the other hand, the supersymmetry properties
of the microscopical configurations have not been examined so far, though the
agreement with the macroscopical description suggests that one should be able to
arrive at the same condition. It would be interesting to check whether this is indeed
the case.


\vspace{1cm}
\noindent
{\bf Acknowledgements}\\

\vsp{-.3cm}
\noindent
We wish to thank Jos Gheerardyn and Jan Rosseel for useful discussions.
The work of B.J. has been done as a Post-doctoral Fellow  of the F.W.O.-Vlaanderen.
B.J. is also partially supported by the European Commission R.T.N.-program
HPRN-CT-2000-00131, by the F.W.O.-Vlaanderen project G0193.00N and by the Belgian Federal
Office for Scientific, Technical and Cultural Affairs through the Interuniversity Attraction
Pole P5/27. The work of Y.L. and D.R-G. has been partially supported by CICYT grant
BFM2003-00313 (Spain). D.R-G. was supported in part by a F.P.U. Fellowship from M.E.C. (Spain).
Y.L. and D.R-G. would like to thank the Institute for Theoretical Physics at the
University of Leuven for its hospitality while part of this work was done.


\end{document}